\documentclass[%
 reprint,
 superscriptaddress,
 amsmath,amssymb,
 aps, pra,
 floatfix,
]{revtex4-2}

\usepackage{graphicx}
\usepackage{dcolumn}
\usepackage{bm}
 
\usepackage{siunitx}
\usepackage{hyperref}
\usepackage{cleveref}
\usepackage{ulem}

\begin{document}
\preprint{APS/123-QED}

\title{THz mixing of high-order harmonics using \texorpdfstring{YBa$_2$Cu$_3$O$_{7-\delta}$}{YBa2Cu3O7-δ} nanobridges}

\author{Núria Alcalde-Herraiz}
\email{herraiz@chalmers.se}
\affiliation{Quantum Device Physics Laboratory, Department of Microtechnology and Nanoscience, Chalmers University of Technology, Gothenburg, Sweden}

\author{Alessia Garibaldi}
\affiliation{Quantum Device Physics Laboratory, Department of Microtechnology and Nanoscience, Chalmers University of Technology, Gothenburg, Sweden}
\affiliation{Department of Clinical Neuroscience, Institute of Neuroscience and Physiology, University of Gothenburg, Gothenburg, Sweden}

\author{Karn Rongrueangkul}
\affiliation{Quantum Device Physics Laboratory, Department of Microtechnology and Nanoscience, Chalmers University of Technology, Gothenburg, Sweden}

\author{Alexei Kalaboukhov}
\affiliation{Quantum Device Physics Laboratory, Department of Microtechnology and Nanoscience, Chalmers University of Technology, Gothenburg, Sweden}

\author{Floriana Lombardi}
\affiliation{Quantum Device Physics Laboratory, Department of Microtechnology and Nanoscience, Chalmers University of Technology, Gothenburg, Sweden}

\author{Sergey Cherednichenko}
\affiliation{Terahertz and Millimeter Wave Laboratory, Department of Microtechnology and Nanoscience, Chalmers University of Technology, Gothenburg, Sweden}

\author{Thilo Bauch}
\email{thilo.bauch@chalmers.se}
\affiliation{Quantum Device Physics Laboratory, Department of Microtechnology and Nanoscience, Chalmers University of Technology, Gothenburg, Sweden}

\date{\today}% It is always \today, today,

\begin{abstract}
Superconducting materials are a key for technologies enabling a large number of devices including THz wave mixers and single photon detectors, though limited at very low temperatures for conventional superconductors. High temperature operation could in principle be offered using cuprate superconductors. However, the complexity of the material in thin film form, the extremely short coherence length and material stability, have hindered the realization of THz devices working at liquid nitrogen temperatures. \texorpdfstring{YBa$_2$Cu$_3$O$_{7-\delta}$}{YBa2Cu3O7-δ} (YBCO) nanodevices have demonstrated non-linear properties typical of Josephson-like behavior, which have the potential for the mixing of AC signals in the THz range due to the large superconducting energy gap. Here, we present AC Josephson functionalities for terahertz waves utilizing Abrikosov vortex motion in nanoscale-confined fully planar YBCO thin film bridges. We observe Shapiro step-like features in the current voltage characteristics when irradiating the device with monochromatic sub-THz waves (\SIrange{100}{215}{\giga\hertz}) at \SI{77}{\kelvin}. We further explore these nonlinear effects by detecting THz high-order harmonic mixing for signals from \SI{200}{\giga\hertz} up to \SI{1.4}{\tera\hertz} using a local oscillator at \SI{100}{\giga\hertz}. Our results open a path to an easy-fabricated HTS nonlinear nanodevice based on dimensional confinement suitable for terahertz applications.
\end{abstract}

\maketitle
Photodetection in the terahertz (THz) frequency range ($f \approx$ \SIrange{0.1}{10}{\tera\hertz})\cite{Rogalski2019, Lewis2019, Qiu2021} is essential in emerging technologies in imaging \cite{li2023}, spectroscopy \cite{Siegel2002} and quantum information \cite{Zmuidzinas2004}. Despite considerable progress in semiconducting devices, superconductors are vital in bridging this frequency range as soon as high sensitivity, high frequency, or large pixel-count imaging are of concern \cite{Day2003,Baselmans2022}.  
Here, typical implementations are transition-edge sensors, hot-electron bolometers, Josephson junctions, kinetic inductance detectors to name a few. However, most implementations rely on low-temperature superconductors (LTS), requiring cryogenic cooling from \SIrange{4}{20}{\kelvin} which significantly limits practical operation.

Implementing devices with critical temperature ($T_C$) above 20 K would alleviate the requirement of costly cryogenic cooling equipment. Here, for example, MgB$_2$ with a $T_C$ slightly below 40~K has been used for THz and IR detectors \cite{Novoselov2017, Charaev2024}. Instead, high-temperature superconductor (HTS) devices, such as those based on  \mbox{YBa$_2$Cu$_3$O$_{7-\delta}$} (YBCO) offer a compelling alternative by enabling in principle operation at liquid nitrogen temperatures achievable with simple and low-cost cryo-coolers. State-of-the-art HTS harmonic mixers employ the non-linear current phase relation (CPR) of Josephson Junctions (JJs) realized using different techniques like bicrystal\cite{Chen1997_PhysC,Tarasov1999}, step-edge\cite{fukumoto1993millimeter,Gao2017}, and oxygen ion-irradiation \cite{Malnou2012, Malnou2014}. However, for these devices, the range of operation frequencies remains restricted to sub-THz frequencies. Higher frequency values (up to 640 GHz) are achieved at temperatures below \SI{20}{\kelvin}\cite{Du2025}. 

While most superconducting mixers rely on barrier-based JJs, there is a growing interest to replace the active element with a simple superconducting constriction that does not contain any barrier. This mitigates fabrication complexity and could avoid detrimental effects due to barrier imperfections \cite{vijay2009optimizing,rieger2023granular}. Indeed, non-linear CPR is not limited to conventional barrier-based Josephson Junctions. Dayem bridges \cite{Anderson1964} offer a scalable route to Josephson-like behavior by nanopatterning a constriction in a bare film \cite{Likharev1979}. Their transport properties, governed by geometrical confinement \cite{likharev1972vortex,Anderson1964}, allow Josephson-like dynamics even when the bridge length exceeds the coherence length ($l  \gg \xi_0$), provided the width is below the Pearl length ($w \ll \lambda_L^2 t^{-1}$ where $t$ is the film thickness and $\lambda_L$ the London penetration depth). In this limit, the Josephson-like behaviour is enabled through coherent Abrikosov vortex dynamics \cite{Likharev1979}. Such structures have explored the physics of non-equilibrium superconductivity \cite{xu2008long,embon2017imaging,llordes2012nanoscale,dobrovolskiy2020ultra, jalabert2023thermalization}
and inspired the development of novel superconducting nanodevices \cite{trabaldo2019grooved,rijckaert2019superconducting,ritter2021superconducting, rocci2020large}.

Here, we demonstrate THz harmonic mixing in YBCO Dayem nanobridges integrated with planar antennas, operating at temperatures up to \SI{77}{\kelvin}. In these devices, the finite voltage state is determined by vortex dynamics. Harmonic mixing up to \SI{1.4}{\tera\hertz} (14th harmonic) is achieved, with the upper frequency limit set by the RF source and measurement set-up rather than by intrinsic device constraints. These findings establish YBCO nanobridges as a promising technology for high-frequency THz detection enabled by coherent Abrikosov dynamics. 
 \begin{figure}
\includegraphics{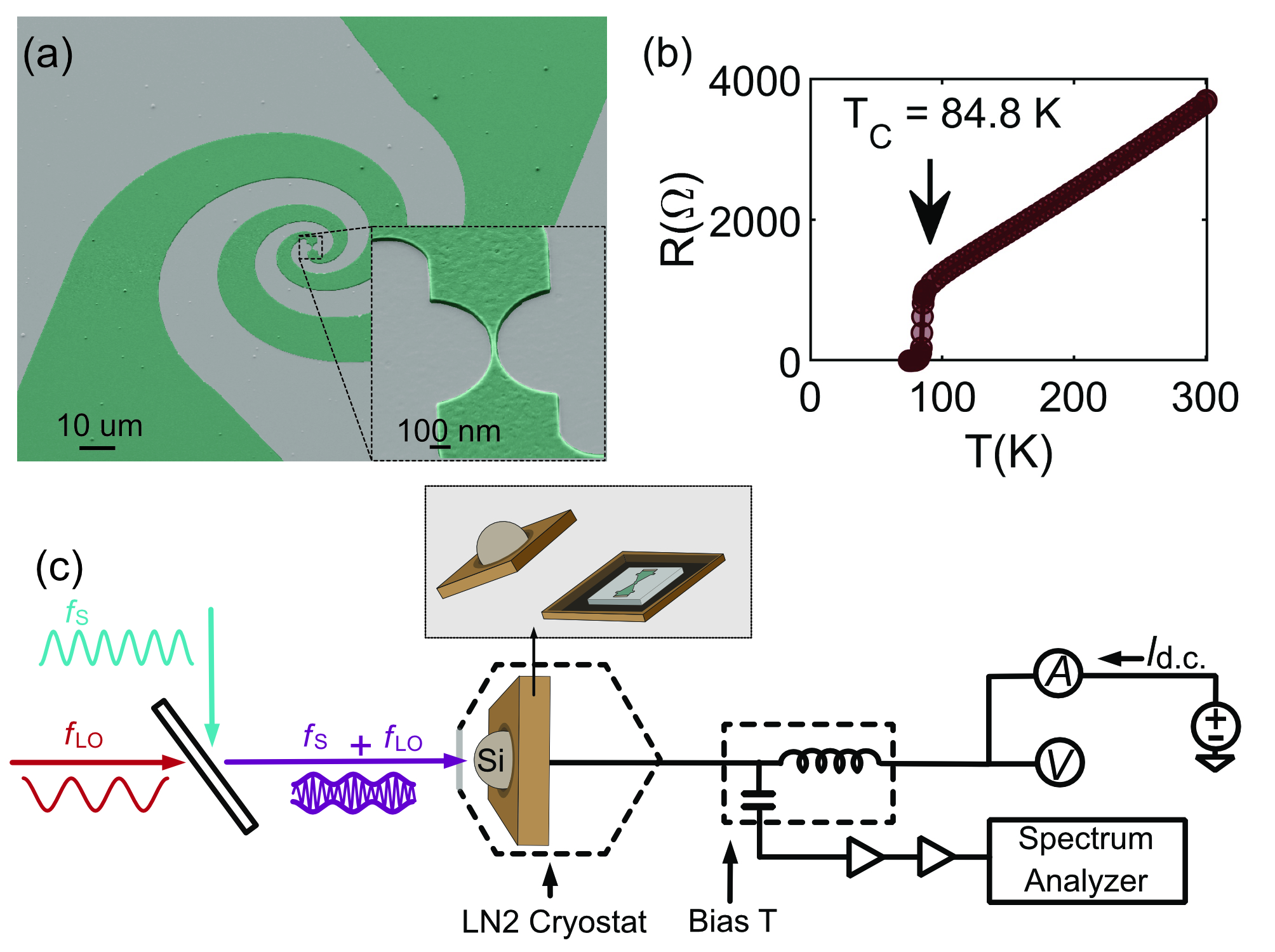}
\caption{\label{fig1} \textbf{Design and characterization of nanobridges heterodyne detectors and measurement set-up.} (a) SEM false-color image of a \SI{70}{\nano\meter} wide YBCO nanowire integrated with a YBCO spiral antenna on a MgO substrate. The inset shows a zoom-in of the nanowire oriented along \textit{a}-axis. (b) Resistance as a function of the temperature of the device. (c) Schematic of the measurement setup with an inset showing the device inside the sample box. }
\end{figure}

The HTS devices were fabricated from \SI{50}{\nano\meter}-thick overdoped, untwinned YBCO films grown on (110)-oriented MgO substrates \cite{Arpaia2019}. The devices layout consists of a YBCO spiral antenna with an embedded bridge as shown in a false-color scanning electron microscope (SEM) image in \Cref{fig1}(a). Each nanobridge  had widths (\textit{w}) and lengths (\textit{l}) of $70 \times 50$  nm$^2$ and $200 \times 400$  nm$^2$, see inset in \Cref{fig1}(a). The devices were patterned by a single-step lithography process combining a carbon mask, electron beam lithography, and Ar ion beam etching \cite{Arpaia2018} (see Supplemental Note 1). This nanofabrication process preserves the superconducting properties of the YBCO film resulting in critical current density approaching the theoretical depairing limit \cite{Nawaz2013PC}. In addition, transmission electron microscope imaging \cite{arpaia2017transport} and noise \cite{trabaldo2020noise} measurements demonstrate that our nanobridges do not contain any grain boundaries. Electrical characterization of our devices was performed in a two-point contact configuration in a DynaCool Physical Properties Measurement System (Quantum Design Inc., USA), yielded a superconducting critical temperature of T$_C$ = \SI{84.8}{\kelvin}, \Cref{fig1}(b), comparable to previously reported overdoped, untwinned \SI{50}{\nano\meter}-thick films \cite{wahlberg2021reshaping}. Furthermore, given the critical current density at \SI{77}{\kelvin} being \SI{1.2e6}{\ampere\per\centi\meter\squared} extrapolation from the Bardeen expression $J_C\propto (1-(T/T_C)^2)^{3/2}$ \cite{bardeen1962critical} would give a value of $\sim$ \SI{2e7}{\ampere\per\centi\meter\squared} at temperatures below 4~K, comparable to our previously reported data\cite{Nawaz2013PC}. 

The devices were characterized by mixing THz signals up to \SI{1.4}{\tera\hertz} in the setup illustrated in \Cref{fig1}(c), see also Supplemental Note 1 for more details. In this configuration it was possible to obtain both the current-voltage characteristics (IVCs) and the intermediate frequency output power of the device as a function of the voltage across the device when using it as a heterodyne mixer. 
%%-------------
\begin{figure}[b]
\includegraphics[scale=1] {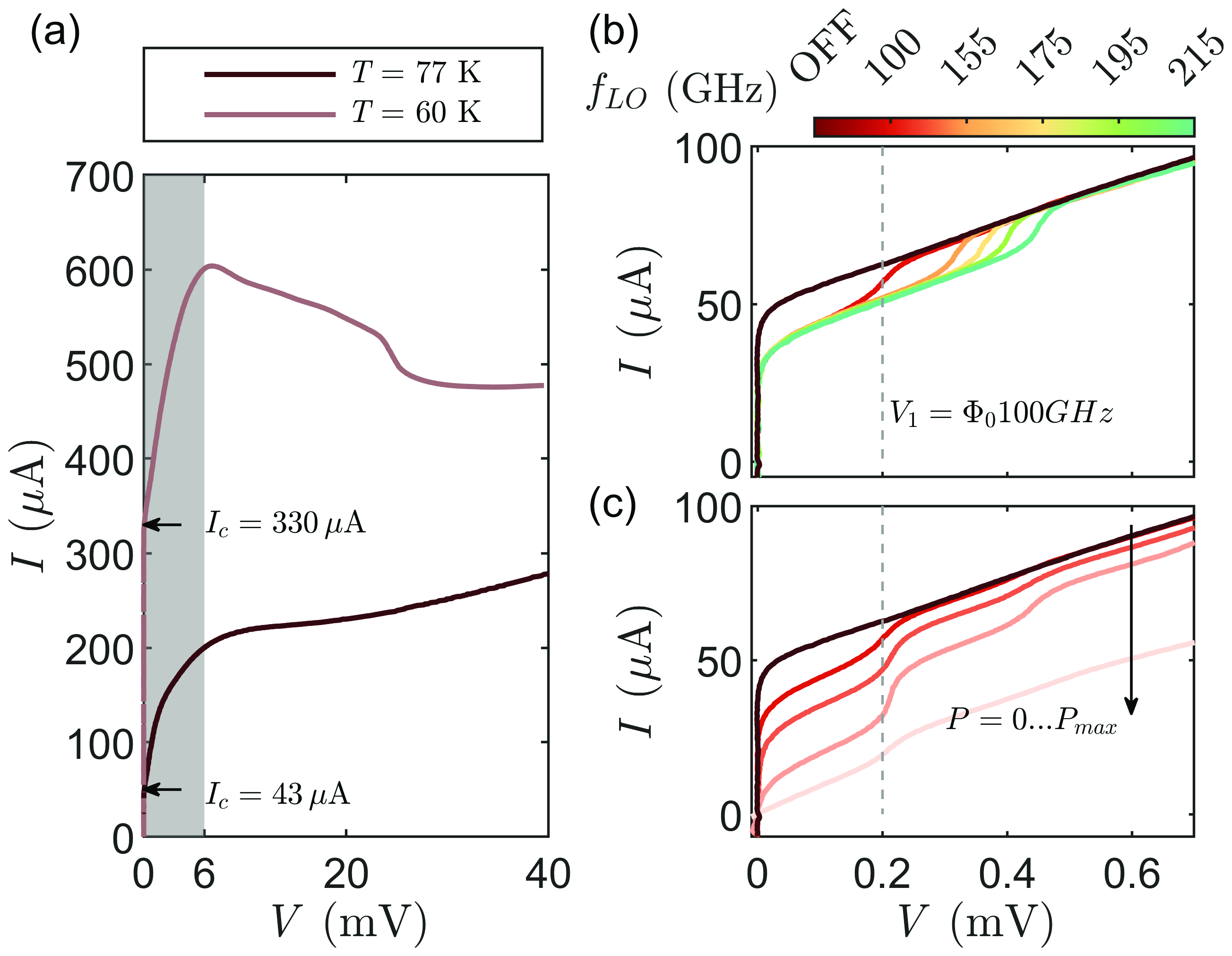}  
\caption{\label{fig2}\textbf{Photo-induced response of a bridge with cross-section of $70 \times 50$  nm$^2$.} (a) IVCs for a nanobridge for temperatures \SI{77}{\kelvin} and \SI{60}{\kelvin} where the critical currents are indicated for both curves (b) Measurement at \SI{77}{\kelvin} of the current-voltage characteristic without an applied signal and with an applied frequency $f_{LO}$ = \SI{100}{\giga\hertz}, \SI{155}{\giga\hertz}, \SI{175}{\giga\hertz}, \SI{195}{\giga\hertz} and \SI{215}{\giga\hertz}.  (c) Evolution at \SI{77}{\kelvin} of the current voltage characteristic for non-illumination case and for increasing illumination power at $f_{LO}$ = \SI{100}{\giga\hertz}.}
\end{figure}

\begin{figure*}
\includegraphics[scale=1]{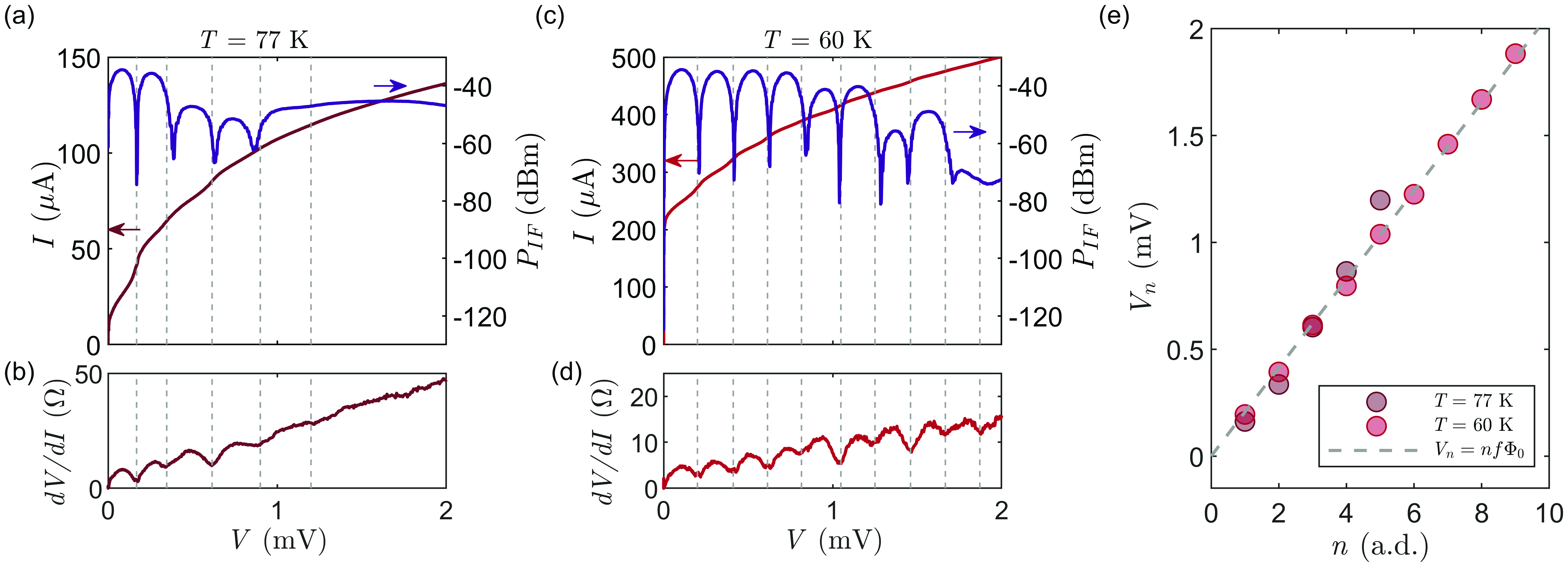}
\caption{\label{fig3}\textbf{Second harmonic mixing using $f_{LO}$ = \SI{100}{\giga\hertz} with $f_{S}$ = \SI{201}{\giga\hertz} for a bridge with cross-section of $70 \times 50$  nm$^2$.} (a) IVC (left axis) and output power response detected at $f_{IF}$ = 1 GHz (right axis) at $T$ = \SI{77}{\kelvin}.  (b) Differential resistance obtained from IVC from panel (a) as a function of voltage. (c) IVCs measurement (left axis) and output power response (right axis) measured at temperature $T$ = \SI{60}{\kelvin}. (d) Differential resistance obtained from panel (c) as a function of voltage. The vertical dashed lines indicate the the voltages corresponding to the minima of differential resistance. (e) Voltage position of current steps from panels (a) and (c) as a function of the step number for $T$ = \SI{77}{\kelvin} and $T$ = \SI{60}{\kelvin}, respectively. The voltage positions are obtained after fitting a Lorentzian to the corresponding minima of the differential resistance. The standard deviation of the fitting is smaller than the marker size.}
\end{figure*}

In the absence of illumination, the IVCs of the nanobridges vary across the studied temperature range, as shown in \Cref{fig2}(a) for the $70 \times 50$  nm$^2$ bridge and which is representative of the wider geometry as discussed in Supplemental Note 2. Increasing the temperature both reduces the critical current and alters the shape of the finite-voltage branch in the IVC. At 60 K, a flux-flow (FF) voltage state, characterized by a downward curvature in the IVC \cite{schneider1993nanobridges} and indicative of vortex dynamics, is observed in the current range 
$I_C < I < \SI{605}{\micro\ampere}$ reaching a maximum voltage of 6 mV (highlighted as grey zone in \Cref{fig2}(a)). After this voltage a flux-flow instability occurs followed by the resistive state with a stable hot-spot (\textit{V}\textgreater\SI{30}{\milli\volt})\cite{skocpol1974self}. It is interesting to note that many studies performed on LTS bridges reveal that the FF regime is only accessible for $T \lesssim T_C $ or under a magnetic field \cite{dobrovolskiy2020ultra, grimaldi2015}. In contrast, in YBCO nanobridges the FF regime expands to broader range of temperatures and does not require an externally applied magnetic field \cite{myoren1989josephson, schneider1993nanobridges, Nawaz2013PC}.

 The nanobridge exhibits Shapiro-like current steps in the FF region which are analogous to the ones observed in conventional JJs\cite{Shapiro1963}, when illuminated with RF signals in the range of \SIrange{100}{215}{\giga \hertz} even at \SI{77}{\kelvin}, see \Cref{fig2}(b). In the case of a bridge, these steps in the IVCs are due to phase-locking the radiation frequency to the coherent motion of Abrikosov vortices and appear at specific voltages $V_n$  given by the relation $V_n = n f_{LO} \Phi_0$ where $V_n$ is the voltage at integer step number $n$, $\Phi_0$ is the magnetic flux quantum, and $f_{LO}$ is the frequency of the external LO, see \Cref{fig2}(b)\cite{likharev1972vortex,aslamazov1975josephson}. Observation of Shapiro steps in the nanobridge required its width to be smaller than the Pearl length $\lambda_P$.  This is the case for our devices considering a film thickness of $t$ = \SI{50}{\nano\meter} and previously reported London penetration depths in similar devices in a range from \SIrange{270}{350}{\nano\meter} \cite{baghdadi2014toward,arpaia2014resistive}. While Shapiro-like steps have been observed in Dayem bridges under microwaves irradiation (\SIrange{1}{20}{\giga \hertz}) at low temperatures (around \SI{4}{\kelvin})\cite{myoren1989josephson,lee1993microwave}, this is the first report of such steps at frequencies above \SI{100}{\giga\hertz} at \SI{77}{\kelvin}. The observation of Shapiro-like steps indicates coherent Abrikosov dynamics in the device which should also enable the detection of harmonic mixing. 

%--------------------------------------------------
\Cref{fig3} shows the second harmonic mixing response of the nanobridge at $T$ = \SI{77}{\kelvin} and \SI{60}{\kelvin}, using a LO at $f_{LO}$ = \SI{100}{\giga\hertz} applied together with a signal $f_{S}$ = \SI{201}{\giga\hertz}. For both temperatures, the LO power was adjusted to suppress the critical current $I_{C}$ to half of its zero-power value to measure at equivalent $I_{AC}/I_{C}$, see Supplemental Note 3. 
\Cref{fig3}(a) (right axis) shows the intermediate frequency (IF) power $P_{IF}$ at $|f_{S} - 2f_{LO}| = 1$ GHz as a function of the DC voltage bias \textit{V} across the nanobridge, with the corresponding IVC (left axis). To resolve current steps at higher voltage bias, we calculated the differential resistance $dV/dI$, see \Cref{fig3}(b). At both temperatures, IF power is measurable up to \SI{2}{\milli\volt}, ranging from \SIrange{-40}{-80}{dBm}. Remarkably, this occurs even at \SI{77}{\kelvin} without any discernible Shapiro steps in the $dV/dI$ at high voltages. 

Vertical dashed lines in \Cref{fig3}(a) and (b) indicate the voltage positions of the minima in $dV/dI$. We first note that at high voltages the minima in $dV/dI$ become offset relative to the minima of the IF output. Also that while the first 3 steps closely follow the expected dependence $V_n = n f_{LO} \Phi_0$, for $n >3$ the spacing between Shapiro-like steps increases with increasing voltage (or $n$), deviating from the expected behaviour. While voltage offsets between the minima in IF output power and Shapiro-step position have also been observed in JJs \cite{Du2017}, the origin of the deviation of these Shapiro-like steps from their expected voltage values remains unclear. Possible mechanisms could be Josephson dissipation or enhanced temperature \cite{Koval2010}. \Cref{fig3}(c) shows that at $T$ = \SI{60}{\kelvin} the current steps in the IVC appear at the expected voltage values, suggesting that the temperature has an effect on the vortex dynamics. This phenomenology is further illustrated in \Cref{fig3}(e), where Lorentzian fits to the $dV/dI$ minima in \Cref{fig3}(b) and \Cref{fig3}(d), yield step voltages $V_n$, when plotted against step number \textit{n} and reveal a deviation from linearity at \SI{77}{\kelvin}, in contrast to the linear behaviour at \SI{60}{\kelvin}. 
%%--------------------------------------------------------------
\begin{figure}%[b]
\includegraphics[scale=1] {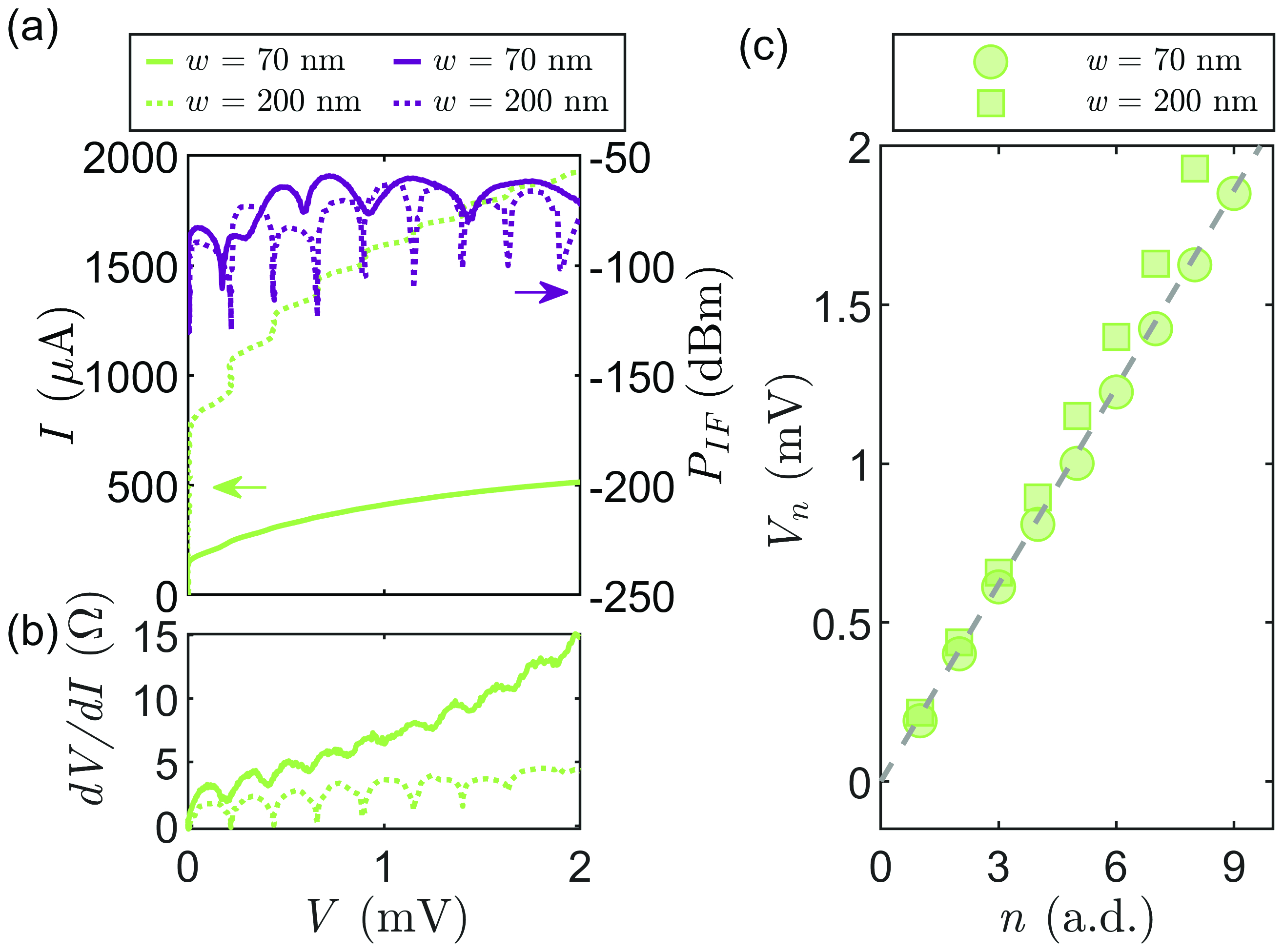}  
\caption{\label{fig4} \textbf{Comparison of tenth harmonic mixing using $f_{LO}$ = \SI{100}{\giga\hertz} and $f_{S}$ = \SI{1001}{\giga\hertz} in bridges with cross-section of $70 \times 50$  nm$^2$ and $200 \times 50$  nm$^2$, biased at equivalent AC currents at temperatures $T$= \SI{59}{\kelvin} and $T$= \SI{58}{\kelvin}, respectively.} (a) IVC measurement (left axis) and output power response detected a $f_{IF}$ = \SI{1}{\giga\hertz} (right axis).  (b) Differential resistance $dV/dI$ obtained from the IVCs shown in (a) as a function of voltage. (c) Voltage positions of current steps obtained from fitting a Lorentzian to the minima of the differential resistances shown in (a) and (c) as a function of the step number. The standard deviation of the fitting is smaller than the marker size.}
\end{figure}

To understand if sub-\SI{100}{\nano\meter} dimensions are necessary for mixing,  we investigate the effect of bridge dimension on the harmonic mixing response. \Cref{fig4} shows the tenth harmonic mixing response for an LO of \SI{100}{\giga\hertz} mixed with a signal frequency $f_{S}$ of \SI{1001}{\giga\hertz} measured in nanobridges with cross-sectional areas of $70 \times 50$  nm$^2$ and $200 \times 50$  nm$^2$ both operated close to \SI{60}{\kelvin}. We observe that the mixing performances are comparable between two devices with less pronounced modulation in $P_{IF}$ for the narrower device. This might offer an operational advantage by broadening the voltage range displaying a flatter power response.

The IF power shown in \Cref{fig4}(a) (right axis) reveals features consistent with those in \Cref{fig3}. The differential resistance (\Cref{fig4}(b)) shows that the $200 \times 50$  nm$^2$ bridge exhibits sharper minima, consistent with better-defined steps in the IVC. Similar to the $70 \times 50$  nm$^2$ bridge at \SI{77}{\kelvin}, the voltage positions $V_n$ against the step number n for wider nanobridge deviate from the linear trend, see \Cref{fig4}(c).  In the wider bridge, this deviation persists even at \SI{60}{\kelvin}, suggesting that temperature further modifies vortex dynamics, affecting Shapiro step positions, but not harmonic mixing. Importantly, this highlights the robustness of high-order harmonic mixing, which remains unaffected by the visibility of Shapiro steps.

%---------------------------------------------------------------
\begin{figure}[b]
\includegraphics[scale=0.09]{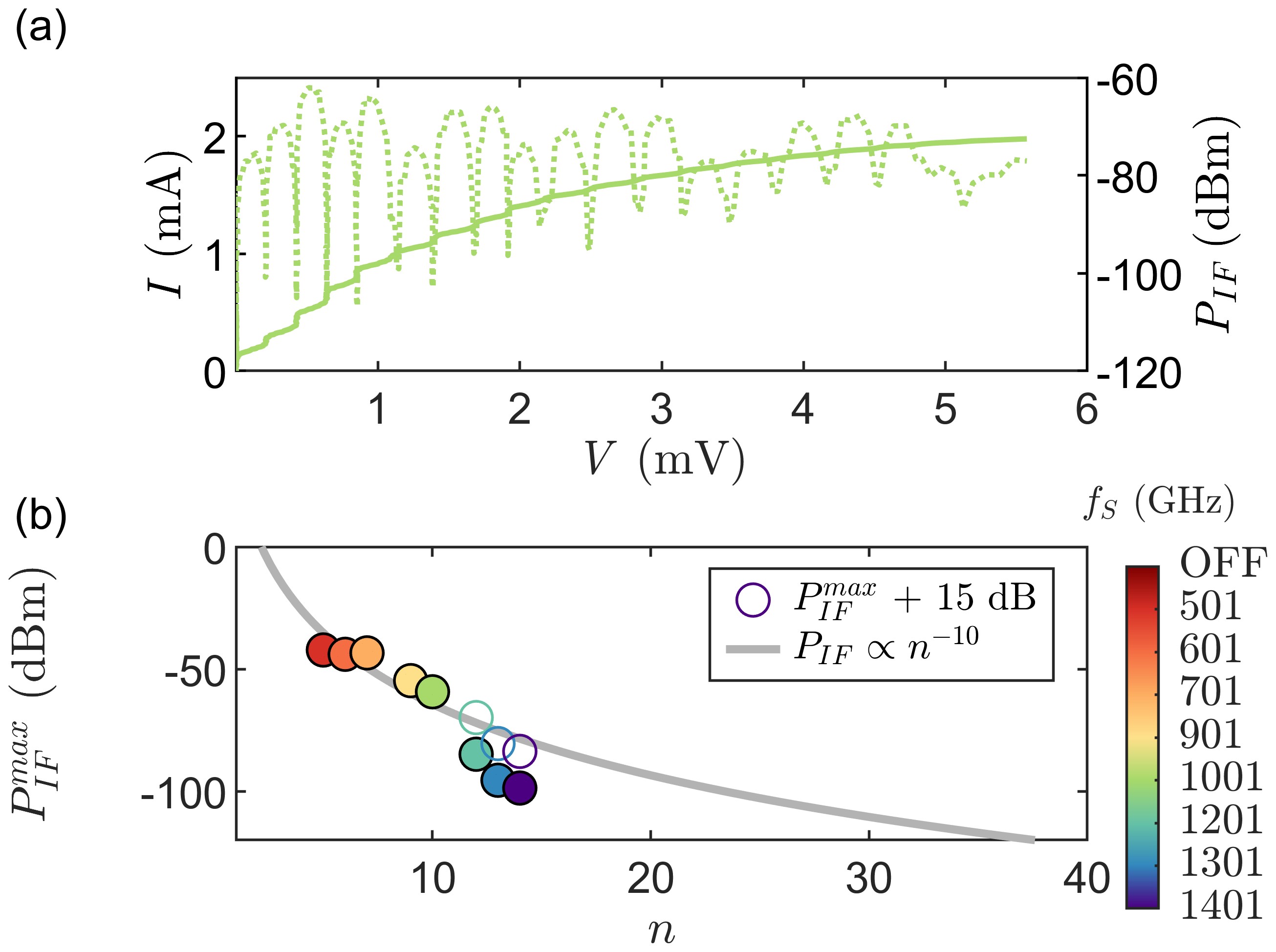}
\caption[Detection limit of the nanobridges at 59-58 K]{%
\label{fig5} 
\textbf{Detection limit of the nanobridges at 59-58 K.} 
(a) For the nanobridge with cross-section of $200 \times 50$ nm$^2$, both \SI{100}{\giga\hertz} current steps (left axis) and harmonic mixing with \SI{1.001}{\tera\hertz} i.e. the tenth harmonic (right axis) extend up to the flux-flow instability voltage (\SI{5.5}{\milli\volt}). 
(b) For the nanobridge of cross-section $70 \times 50$ nm$^2$, intermediate frequency power as a function of the harmonic number n (filled circles) for a constant LO frequency (\SI{100}{\giga\hertz}). Open circles are corrected for a 15 dB (x30) lower THz source power in the 1.1-1.5 THz range (x12, x13, x14 harmonics).  
The solid line is a $P_{IF} \propto n^{-10}$ fit. The y-axis is limited to -120 dBm which corresponds to our noise threshold.}
\end{figure}

It is worth discussing the ultimate limit of harmonic mixing in YBCO nanobridge detectors which is instrumental for device implementation. In this work, we identify two key factors limiting the mixing response: the superconducting energy gap and vortex dynamics.

A fundamental constraint is set by the superconducting energy gap of YBCO, which defines the upper voltage, and thus frequency limit for mixer operation. At \SI{60}{\kelvin}, the gap corresponds to approximately \SI{12.5}{\milli\volt} \cite{wang2020portable}, which would typically suggest an upper frequency limit of around \SI{6}{\tera\hertz} for the AC Josephson effect. However, experiments on Nb Josephson junctions have shown that, especially at low temperatures, Shapiro steps and harmonic mixing can still be observed at frequencies exceeding the superconducting gap value \cite{mcdonald1972four}.

A second intrinsic limitation arises from vortex dynamics in the nanobridge. As shown  in Figure 2(a), vortex dynamics should be present through the flux-flow branch extending up to \SI{6}{\milli\volt}, beyond which we observe a flux-flow instability. This behavior is further supported by \Cref{fig5}(a), which shows the characteristic minima in the IF power response associated to Shapiro-like steps in the IVC up to $5.5~$mV, indicating that vortex motion remains phase-locked to the LO even at high voltages.  The onset of flux-flow instability beyond this voltage is dependent on the maximum vortex velocity achievable in our material, which is influenced by both electron diffusivity and quasiparticle relaxation rate \cite{lee1993microwave,xu2008long,bezuglyj2019local}.
Whether the \SI{6}{\milli\volt} threshold sets a fundamental limit on the maximum signal frequency remains an open question.  More studies on this phenomenology would be valuable not only for optimizing THz detection in the current devices but also for other applications where fast vortex motion is of interest.  
In principle, one could even extend the flux-flow region using thinner films since our preliminary investigation shows the flux-flow regime extends up to 50 mV at 4~K in optimally doped samples \cite{wahlberg2021reshaping}. Therefore, thickness and doping dependent mixing response will be instrumental in understanding if one can push the vortex flow regime to even higher voltages.

In this study of high-order harmonic mixing in YBCO Dayem bridges using frequency locking to the vortex motion we utilized a single LO source (100 GHz) with a test signal varying in a broad spectral range, \SIrange{200}{1400}{\giga\hertz}. Although focusing on a more specific THz range would allow for a thorough performance optimization, we attempt to obtain a general understanding of the device operation as a coherent THz detector. In particular, we focused on both reaching high THz frequencies and high harmonic order. Previous studies on both YBCO JJs\cite{Chen1997_PhysC,Du2017,fukumoto1993millimeter} and equivalent Schottky diode\cite{jayasankar2021} harmonic mixers indicate a reduction of the mixing efficiency as the order of the harmonic increases following the empirical trend $P_{IF} \propto n^{\alpha}$ that reflects on the conversion efficiency $\eta = P_{IF}/P_{RF}$. From our measurements, we extract the averaged $P^{max}_{IF}$ which is plotted against harmonic number $n$, correcting the source power variations at higher harmonics indicated with open circles in \Cref{fig5}(b). The resulting decay yields $\alpha$= -10. This value is an overestimation as increasing THz signal frequency, increases losses in the YBCO-spiral antenna (including variation of its elliptical polarization), optical coupling loss, etc. However, with the current device, extrapolation based on our setup suggests a detection limit of up to \SI{3.5}{\tera\hertz} (see \Cref{fig5}(b)). 
Therefore, at this stage the presented devices already demonstrate performance levels suitable for many practical applications, such as spectroscopy\cite{bidgoli2014, richter2024} or wireless communication\cite{federici2010}. Moreover, beyond their current technical capabilities, these devices offer a promising platform for exploring vortex dynamics potentially approaching the superconducting gap frequency.

\section*{\textbf{Data availability }} 
The data that support the findings of this study are available from the corresponding author upon reasonable request. 
 
\begin{acknowledgments}
We acknowledge support by the Swedish Research Council (VR), under the Projects 2022–04334 (F.L.), 2020-05184 (T.B), 2024-05138 (S.C.). We also acknowledge support from the Swedish infrastructure for micro- and nanofabrication - MyFab, Area of Advance Nano and Information and Communication programs at Chalmers University of Technology. 
\end{acknowledgments}

\bibliography{Maintext_NAH_Oct2025}
\end{document}